\documentstyle[aps,12pt,preprint]{revtex}

\def\pmb#1{\setbox0=\hbox{$#1$}%
\kern-.025em\copy0\kern-\wd0
\kern.05em\copy0\kern-\wd0
\kern-.025em\raise.0433em\box0}

\let\varkappa\kappa
\skip\footins = 14pt 
\newif\ifMarginNotes \MarginNotesfalse

\def\mrgn#1{\ifMarginNotes\setbox0=\vtop{\hsize 10pc
   {\footnotesize\noindent\relax #1\par}}\leavevmode
   \vadjust{\dimen0=\dp0 \dimen1=\ht0\advance\dimen1 by .5ex
  \advance\dimen0 by -.5ex
  \kern-\dimen1\hbox{\kern\hsize\kern.5pc$\leftarrow$
  \box0}\kern-\dimen0}\fi}

\ifMarginNotes
\global\def\Mrgnpar#1{\marginpar[\raggedright\footnotesize
    $\rightarrow$\enspace#1]{\raggedright\footnotesize$\leftarrow$\enspace#1}}%
\global\def\Mrgnparb#1{\marginpar[\raggedright\footnotesize#1\enspace
    $\searrow$]{\raggedright\footnotesize$\swarrow$\enspace#1}}%
\else
\global\def\Mrgnpar#1{\relax}%
\global\def\Mrgnparb#1{\relax}%
\fi

\begin{document}
\tighten
\baselineskip=2\baselineskip
\nonfrenchspacing
\flushbottom
\title{Charged Particles in a 2+1 Curved Background\footnotemark[1]}

\footnotetext[1] {\baselineskip=16pt This work is supported in part by funds
provided by  the U.S.~Department of Energy (D.O.E.) under contracts
\#DE-FC02-94ER40818, \#DE-FG02-91ER40676 and, in part, by INFN.  
\hfil \break 
\hfil MIT-CTP-2700
\hfil December, 1997 -- revised March, 1998
\break}

\author{P. Maraner}
\bigskip

\address{Center for Theoretical Physics, Laboratory for Nuclear Science and
INFN, \\ Massachusetts Institute of Technology, Cambridge, MA~02139--4307, USA}

\maketitle

\vspace{1.5in}

\begin{abstract}
The coupling to a 2+1 background geometry of a quantized charged 
test particle in a strong magnetic field is analyzed. 
Canonical operators adapting to the fast and slow 
freedoms produce a natural expansion 
in the inverse square root of the magnetic field
strength. The fast freedom is solved to the second order. 
 At any given time, space is parameterized by a couple 
of conjugate operators and effectively behaves as 
the `phase space' of the slow freedom. The slow Hamiltonian 
depends on the magnetic field norm, its covariant derivatives, 
the scalar curvature and presents a peculiar coupling with the 
spin-connection.
\end{abstract}

%

\setcounter{page}{0}
\thispagestyle{empty}

\newpage
\section{Introduction}

 The dynamics of a charged particle in a given electromagnetic and 
gravitational background is an important problem having implications
in several areas of theoretical and mathematical physics --from 
classical gravity to condensed matter and plasma physics to quantum
field theory. 
 As a quite interdisciplinary example, it represents
the first step to take in addressing the study of a plasma around 
a compact astrophysical object or, more in general, in space and 
cosmological phenomena \cite{ast}. 
 Exact solutions are found when metric and electromagnetic 
two-form share common symmetries. Various special cases  
have been worked out --expecially in two spatial dimensions-- 
with particular emphasis on the underlying algebraic and 
analytical structures \cite{Dun}.   
 Beyond symmetry --in spite of the apparent simplicity--
the general problem displays an extreme degree of complication. 
Classical motion is generally chaotic and one has to be content
with approximate analysis. Even like this, however, the task to 
set down an appropriate perturbative expansion is not strait forward.    
The peculiar structure of the electromagnetic interaction
makes ordinary Hamiltonian perturbation theory not directly 
applicable \cite{LL}. In the fifties and sixties, the 
urgency of the problem in classical 
applications  --expecially in connection with the investigation 
of earth's magnetosphere and in the design of mirror machines for 
the confinement of hot plasma--
motivated Bogoliubov, Kruskal and others to formulate 
adiabatic perturbation theory directly in terms of the equations 
of motion. This led Northrop and Teller to the familiar `guiding
center' picture of the effective dynamics of a charged particle 
in an inhomogeneous magnetic field in a flat spacetime \cite{gca}.   
 Modern applications --ranging from  geodesics motion around 
charged black holes in classical gravity to two-dimensional system 
of non-relativistic electrons in quantum-Hall like devices to plasma 
in astrophysics and cosmology to the investigation of the coupling 
with matter fields in toy models for quantum gravity--
deal more in general with curved backgrounds 
and require the extension of the perturbative analysis 
developed in classical physics to the quantum-mechanical and 
filed-theoretical context. 
 To this task a whole canonical approach to the problem has to 
be developed. This is the aim of the present investigation.

In this paper we address the subject by discussing the effective 
motion of a charged particle in a 2+1 curved background. This 
allows to display better the peculiar canonical structure of 
the system, avoiding complications arising from extra dimensions.
Moreover, the restriction is not just a mathematical artefact. 
The solution in two spatial dimensions is indeed a key ingredient
in the discussion of the relativistic four-dimensional, as well 
as the non-relativistic three-dimensional, cases.  
 From a rather different viewpoint the problem is also
equivalent to the investigation of the effective dynamics 
of a test particle experiencing the `geometric gravitational'
force of Cangemi and Jackiw in a Wick rotated two-dimensional
space-time \cite{CJ}.

 Our analysis is based on the canonical structure of the system and   
is essential the same for the classical and the quantum case.
For definiteness we consider the quantum case. The classical limit
may be obtained straightforwardly. The topology of space-time is 
supposed to be trivial --the direct product of a surface
$\Sigma$ diffeomorphic to the plane and time-- so that all
the local quantities get automatically global definition; 
eg. Ricci rotation coefficients define a spin-connection.
Under these hypothesis it is always possible to choose coordinates
in such a way that the metric takes the form: $g_{00}=1$, 
$g_{0\mu}=0$, $g_{\mu\nu}=$ arbitrary functions of time and
spatial coordinates; $\mu,\nu=1,2$. 
We also assume the electromagnetic field to be purely magnetic. 
Both relativistic and non-relativistic 
problems reduce then to the study of the Hamiltonian of a 
charged particle on the curved surface $\Sigma$.

The paper is organized as follows. In section II we discuss the 
canonical structure of the problem showing how the strong magnetic  
regime naturally produces an expansion in the inverse square root
of the field strength. In section III an adapted set of canonical
operators is introduced. This allows to separate the fast freedom from
the slow one, identifying the adiabatic invariant of the system.   
 The coupling with  background geometry is studied in section 
IV. Besides contributions depending on the scalar curvature and
on covariant derivatives of the magnetic field norm we find a peculiar 
coupling with the spin-connection. The theory is general as
well as `Lorenz' covariant. Our main result is the effective 
Hamiltonian (\ref{Heff}).
The example of a particle on a conical surface in an axisymmetric 
magnetic field decreasing as the inverse of the distance from the 
vertex is presented in section V.
The last section contains our conclusions. 
In the appendix the necessary technology for maximally simplifying 
the study of the adiabatic expansion is summarized.

\newpage
\section{Charged particle on a curved surface}

We consider a charged scalar particle on a two dimensional surface 
$\Sigma$ in a {\em strong} magnetic background. The surface is 
parameterized by arbitrary coordinates $x^\mu$, $\mu=1,2$, and its
geometry is given by the metric tensor $g_{\mu\nu}$.
The magnetic field is described by a closed antisymmetric two-form 
$b_{\mu\nu}$. 
 In both the non-relativistic and relativistic cases the discussion of
the dynamical problem reduces to studying the Hamiltonian
\begin{equation}
{\cal H}={1\over2}g^{-1/2}\Pi_\mu g^{\mu\nu}g^{1/2}\Pi_\nu
\label{H1}
\end{equation}
The kinematical momenta $\Pi_\mu=-i\partial_\mu-{l_B}^{-2}a_\mu$
have been introduced, $[\Pi_\mu,\Pi_\nu]=ib_{\mu\nu}(\vec{x})$ 
and the physical dimension of the field
is re-adsorbed in the scale factor $l_B$. 
The wavefunction of the system is normalized with respect to the 
measure $\sqrt{g}dx^1dx^2$.
Our analysis is based on the smallness of the magnetic length $l_B$. 
 Throughout our discussion we assume the background scalar curvature 
$R$ as well as the derivatives of the magnetic field norm 
$b=\sqrt{b_{\mu\nu}b^{\mu\nu}/2}$ to satisfy the 
conditions: $|R|\ll{l_B}^{-2}$, $|\triangle b/b|\ll{l_B}^{-2}$ 
and $|\nabla b/b|\ll{l_B}^{-1}$.

 First of all, we focus on kinematics. In absence 
of magnetic interaction the essential operators appearing in the 
description of the system are the coordinates $x^\mu$ 
and the derivatives $-i\partial_\mu$.
These appear as a couple of conjugate variables, 
$[x^\mu,-i\partial_\mu]=i\delta^\mu_\nu$. 
Introducing the magnetic interaction replaces the $-i\partial_\mu$s by the 
non-commuting $\Pi_\mu$s. 
In other words the magnetic background produces a twist 
of the canonical structure.
This is made explicit by transforming to  Darboux coordinate frames
$\xi^{\mu}=\xi^{\mu}(x)$ in which the magnetic field strength takes
the form 
\begin{equation}
b_{\mu\nu}(\xi)={l_B}^{-2}\varepsilon_{\mu\nu}
\end{equation}
($\varepsilon_{\mu\nu}$ is the completely antisymmetric tensor in two
dimensions). Darboux theorem ensure the existence of a well defined 
atlas of such frames. In the new frames $\Pi_1$
and $\Pi_2$ appear as reciprocally conjugate while their commutators 
with the coordinates are still different from zero. On the other hand,
$[\xi^1,\Pi_1]$ and $[\xi^2,\Pi_2]$ are order ${l_B}^2$ compared to 
$[\Pi_1,\Pi_2]$. This make clear that in the strong magnetic  
regime it is convenient to abandon  the description in terms of 
the $\xi^\mu$s and $-i\partial_\mu$s introducing besides $\Pi_1$ and
$\Pi_2$ a new couple of canonical variables. These turns out to be 
the guiding center operators $\Xi^\mu=\xi^\mu+{l_B}^2
\varepsilon^{\mu\nu}\Pi_\nu$ (in this paper we adopt the notation 
$\varepsilon^{\mu\nu}=\varepsilon_{\mu\nu}$). 
Rescaling for convenience the 
$\Pi_\mu$s by $\Pi_\mu\rightarrow{l_B}\Pi_\mu$ --and hence the magnetic
field norm $b$ and the Hamiltonian ${\cal H}$ by a factor ${l_B}^2$-- 
the fundamental commutation relation may finally be re-casted in the form
\begin{equation}
[\Pi_1,\Pi_2]=i\ \ \ \ \ [\Xi^2,\Xi^1]=i{\l_B}^2
\label{ccr}
\end{equation}  
The presence of the small parameter ${l_B}$ in the second relation 
displays the guiding center operators as slow variables of the system.
The physical interpretation of the new quantities emerges by
considering dynamics in the semiclassical regime \cite{Lit,Mar}: 
the $\Pi_\mu$s
take into account the rapid rotation of the particle while the $\Xi^\mu$s
the slow drift of the center of the orbit -the guiding center- on the 
surface.

 Outlined the peculiar canonical structure we come back
to the dynamical problem. This is in general of a certain 
complication\footnotemark[2] the two freedoms of the system being coupled 
by the metric background $g_{\mu\nu}$ as well as by the magnetic 
field strength $b_{\mu\nu}$. 
\footnotetext[2] {\baselineskip=16pt An exact solution of the problem
is only possible when metric and magnetic field share a common 
symmetry. Typical examples are the Landau problem --motion on a plane
in a uniform magnetic field-- the motion on a sphere in a monopole
field and the motion on the Poincar\'e half plane in a hyperbolic magnetic 
field \cite{Dun}. The exact (degenerate) ground state of the system 
may be obtained whenever metric and magnetic two-form define a K\"ahler
structure on the surface $\Sigma$ \cite{KO}. More general conditions 
require an approximate analysis.}
Note that even starting from a simple
geometrical context --eg. a flat one-- transforming to Darboux 
frames produces a quite complicated form of the interaction.
 Nevertheless, whenever the curvature radii of the surface
and the variation length scale of the magnetic field may be considered
larger than the magnetic length ${l_B}$, it turns out  
possible to perform an approximate analysis in quite general terms. 
We start, of course, from 
(\ref{H1}). As a first technical steep we rescale wavefunction 
and Hamiltonian by $\psi\rightarrow g^{1/4}\psi$ and ${\cal H}
\rightarrow g^{1/4}{\cal H}g^{-1/4}$. This changes the integration
measure from $\sqrt{g}dx^1dx^2$ to $dx^1dx^2$ making the Hamiltonian
more symmetric and simplifying further manipulations. The second steep
is that of adapting variables. We transform therefore to  
Darboux coordinate frames according to the above kinematical
discussion.
The transformed metric tensor is denoted by $\gamma_{\mu\nu}$. 
Observe that in these preferential frames the metric 
determinant $\gamma$ is related to the magnetic field norm  by
$\gamma=b^{-2}$. Moreover, all the functions of the
coordinates have now to be evaluated in 
$\xi^\mu=\Xi^\mu-l_B\varepsilon^{\mu\nu}\Pi_\nu$
producing a natural expansion of the Hamiltonian in the small
parameter $l_B$.
 Taking into account the rescaling of the $\Pi_\mu$s, the one of
wavefunction and Hamiltonian and expanding in $l_B$, 
(\ref{H1}) takes the form
\begin{eqnarray}
{l_B}^2{\cal H}&=& {1\over2}\gamma^{\mu\nu}\Pi_\mu\Pi_\nu
         -{l_B\over2}(\partial_\kappa\gamma^{\mu\nu})
          \varepsilon^{\kappa\rho}\Pi_\mu\Pi_\rho\Pi_\nu +\nonumber\\
        & &+{{l_B}^2\over4}(\partial_\kappa\partial_\lambda\gamma^{\mu\nu})
          \varepsilon^{\kappa\rho}\varepsilon^{\lambda\sigma}
          \Pi_\mu\Pi_\rho\Pi_\sigma\Pi_\nu
        -{{l_B}^2\over4}{\triangle b\over b}
         +{{l_B}^2\over8}{|\nabla b|^2\over b^2}+{\cal O}({l_B}^3)
\label{H2}
\end{eqnarray}
where the inverse metric $\gamma^{\mu\nu}$, the magnetic field norm
$b$ and all their derivatives are evaluated in the guiding center
operators $\Xi^\mu$.

\section{Spinning and drifting}

 We first focus on the zero order of expansion (\ref{H2}) by discussing
the truncated Hamiltonian 
 ${\cal H}^{(0)}={1\over2}\gamma^{\mu\nu}\Pi_\mu\Pi_\nu$.
This is quadratic in the $\Pi_\mu$s with coefficients depending
on the slow variables $\Xi^\mu$. It should therefore be  possible to reduce 
the problem to an harmonic oscillator up to higher orders in $l_B$. 
To this task we consider the decomposition of $\gamma^{\mu\nu}$ 
in terms of the zwei-beinen ${e_i}^\mu$; $\gamma^{\mu\nu}={e_i}^\mu{e_i}^\nu$.
We then introduce the `normalized zwei-beinen'
${n_i}^\mu=b^{-1/2}{e_i}^\mu$ in such a way that 
\begin{equation}
{\cal H}^{(0)}={1\over2}\left[{n_i}^\mu(\Xi)\Pi_\mu\right] 
                        b(\Xi)
                        \left[{n_i}^\nu(\Xi)\Pi_\nu\right]
-{i\over4}{n_i}^\mu(\Xi) b(\Xi){n_i}^\nu(\Xi)\varepsilon_{\mu\nu}
\label{H01}
\end{equation}
It is clear that the new $\bar{\Pi}_\mu$s recasting ${\cal H}^{(0)}$ 
in an harmonic oscillator Hamiltonian should have the form 
$\bar{\Pi}_i={n_i}^\mu(\Xi)\Pi_\mu+{\cal O}({l_B}^2)$.
To obtain a genuine set of canonical variables
--not a perturbative one-- we produce the rotation of the $\Pi_\mu$s
in the ${n_i}^\mu$ directions by mean of the unitary transformation 
\begin{equation}
U=\exp\left\{-{i\over4}\varepsilon^{\mu i}{[\lg n]_i}^\nu
                              \{\Pi_\mu,\Pi_\nu\}\right\} 
\end{equation}
The new canonical operators are  defined by 
$\bar{\Pi}_i=\delta_i^\mu U\Pi_\mu U^{\dagger}$ and 
$\bar{\Xi}^\mu=U\Xi^\mu U^{\dagger}$.
An explicit expression as a power series in ${l_B}^2$ may now be 
obtained to any order. As a preparation for the next
section we write the new variables to the order ${l_B}^2$. Introducing
the non-covariant rotation coefficients 
${\rho_i}^j,_k={n_k}^\mu(\partial_\mu{n_i}^\nu){n_\nu}^j$
we have
\begin{eqnarray}
\bar{\Pi}_i&=&{n_i}^\mu\Pi_\mu-
            {{l_B}^2\over8}\varepsilon^{mn}
            {\rho_i}^k,_m {\rho_j}^l,_n
            \varepsilon^{jh} {n_k}^\kappa{n_h}^\mu{n_l}^\lambda 
            (\Pi_\kappa\Pi_\mu\Pi_\lambda+\Pi_\lambda\Pi_\mu\Pi_\kappa)
            +{\cal O}({l_B}^4) \label{P} \\
\bar{\Xi}^\mu&=&\Xi^\mu-
            {{l_B}^2\over4}\varepsilon^{mn}
            {n_m}^\mu {\rho_j}^l,_n
            \varepsilon^{jk} {n_k}^\kappa{n_l}^\lambda
            (\Pi_\kappa\Pi_\lambda+\Pi_\kappa\Pi_\lambda) 
            +{\cal O}({l_B}^4) \label{X}
\end{eqnarray}
where all the functions on the right hand side are evaluated in $\Xi$.
In order to rewrite (\ref{H01}) in terms of the new operators these 
relations have to be inverted. The task is straightforward yielding 
\begin{equation}
{n_i}^\mu(\Xi)\Pi_\mu=\bar{\Pi}_i+
            {{l_B}^2\over8}\varepsilon^{mn}
            {\rho_i}^k,_m {\rho_j}^l,_n
            \varepsilon^{jh}
            (\bar{\Pi}_k\bar{\Pi}_h\bar{\Pi}_l
            +\bar{\Pi}_l\bar{\Pi}_h\bar{\Pi}_k)+{\cal O}({l_B}^4)
\label{nP}
\end{equation}
and 
\begin{equation}
\Xi^\mu=\bar{\Xi}^\mu+
        {{l_B}^2\over4}
        \varepsilon^{mn}{n_m}^\mu {\rho_j}^l,_n
        \varepsilon^{jk}(\bar{\Pi}_k\bar{\Pi}_l+\bar{\Pi}_l\bar{\Pi}_k)
        +{\cal O}({l_B}^4)
\label{nX}
\end{equation}
in both equations the functions on the right hand side  are 
now evaluated in $\bar{\Xi}$. The substitution of (\ref{nP}) and 
(\ref{nX}) in (\ref{H01}) produces the zero order Hamiltonian as 
a power series in ${l_B}^2$. Introducing the harmonic oscillator 
$J={1\over2}({\bar{\Pi}_1}^2+{\bar{\Pi}_2}^2)$ we obtain 
\begin{equation}
{\cal H}^{(0)}= b(\bar{\Xi}) J 
             + {l_B}^2{\cal H}^{(0,2)}(\bar{\Xi},\bar{\Pi})
             + {\cal O}({l_B}^4) 
\label{H02}
\end{equation}
where ${\cal H}^{(0,2)}$ is a quite complicated expression --quartic 
in the $\bar{\Pi}_\mu$s and depending on the $\bar{\Xi}^\mu$s through
$b$ and the ${\rho_i}^j,_k$s-- which may be evaluated by direct substitution.

 The adiabatic behavior of the system in the strong magnetic regime
may now be read in the first term of expansion (\ref{H02}). 
The fast and slow freedoms decouple up higher order in $l_B$.
The fast freedom is frozen in one of the harmonic 
oscillator eigenstates of the adiabatic invariant $J$. While 
`spinning', the particle drifts on the surface $\Sigma$. 
The drifting is Hamiltonian: the configuration space $\Sigma$ 
appears now as the phase space of the slow freedom; the magnetic 
field norm $b(\bar{\Xi}^1,\bar{\Xi}^2)$ evaluated in the 
couple of conjugate variables $\bar{\Xi}^1$ and $\bar{\Xi}^2$ 
is the Hamiltonian operator governing the slow motion
(see ref.\cite{DJ}).
 
 The situation is substantially analog to the motion on plane
\cite{Wit,Mar}, the metric appearing only in the evaluation of the 
magnetic field norm. The crucial difference  
is that in a non trivial geometrical 
background a constant value of $b$ does not produce in general
the slow variables as exact constants of motion.

\section{Coupling to background geometry}

 We now study the higher order corrections to the effective motion of 
the charged particle. To this task we proceed by the so called
averaging method (see appendix) that is by performing a series of
near-identity unitary transformations separating, order by order 
in $l_B$, the fast freedom from the slow one. 
First of all a little preparation is necessary. 

We re-express all the quantities appearing in the expansion (\ref{H2})
in terms of the new canonical variables $\bar{\Pi}_i$
and $\bar{\Xi}^\mu$. This produces the replacements 
of all the curved space indices $\mu,\nu, ...$ by the flat space indices
$i,k,...$\ . Every `general covariant' index $\mu$ is replaced by
a `Galilei covariant' index $i$ according to the usual rules 
$v_i={e_i}^\mu v_\mu$, $v^i={e^i}_\mu v^\mu$ etc.\ . 

As a second steep it is useful to work out a few basic geometrical 
identities holding in every Darboux frame. These will be precious 
in bringing the adiabatic expansion in an explicit covariant form.
By derivating the relation between metric determinant and magnetic field norm
we obtain: 
$(\partial_\rho\gamma^{\mu\kappa})\gamma^{\nu\lambda}
                                      \varepsilon_{\kappa\lambda}
-(\partial_\rho\gamma^{\nu\kappa})\gamma^{\mu\lambda}
                                      \varepsilon_{\kappa\lambda}=
2b(\partial_\rho b)\varepsilon^{\mu\nu}$. 
Contracting with $\varepsilon_{\mu\nu}$ and rewriting in terms of 
flat space indices yields
\begin{equation}
\Gamma_{ij}^j=-b^{-1}(\partial_ib)
\label{r1}
\end{equation}
(which is the usual relation $\Gamma_{\mu\nu}^\nu=\partial_\mu\lg{g^{1/2}}$
evaluated in a Darboux frame). By multiplying the relation by itself,
contracting and rewriting in terms of flat space indices we also obtain 
\begin{equation}
  \Gamma_{ij}^k\Gamma_{ij}^k
- \Gamma_{ij}^k\Gamma_{ik}^j
- \Gamma_{ij}^j\Gamma_{ik}^k
+2\Gamma_{ii}^j\Gamma_{jk}^k
-\Gamma_{ii}^k\Gamma_{jj}^k=0
\label{r2}
\end{equation}
No other general relations hold among the various contractions
of the Christoffel symbols. 

 We proceed now by evaluating the contributions produced by 
the zero, first and second order terms of (\ref{H2}). 
Everywhere in what follows equation (\ref{r2}) is used to eliminate
$\Gamma_{ii}^k\Gamma_{jj}^k$ in favor of the other four possible 
contractions of the Christoffel symbols.

\subsection*{Second order contribution from ${\cal H}^{(0)}$}

We start the averaging procedure considering the second order
contribution produced by ${\cal H}^{(0,2)}$. To this task it
is necessary to re-express the non-geometrical quantities ${\rho_i}^j,_k$
in terms of the spin-connection 
$\omega_{ij},_k=(\nabla_{e_k}e_i)\cdot e_j$
and the Christoffel symbols $\Gamma_{ij}^k$. A quick computation
yields
\begin{equation}
b^{1/2}{\rho_i}^j,_k= {\omega_i}^j,_k
                    +{1\over2}\delta^i_j\Gamma_{kl}^l
                    -\Gamma_{ik}^j
\label{rsg}
\end{equation}
We recall that the spin-connection is completely antisymmetric in the 
indices $i$ and $j$. In two dimensions it may therefore be rewritten 
in terms of a $U(1)$ gauge potential as  $\omega_{ij},_k=
\omega_k\varepsilon_{ij}$. A point dependent rotation by an angle
$\chi(\xi)$ of the zwei-beinen ${e_i}^\mu$ produces 
the gauge transformation $\omega_k\rightarrow\omega_k+\partial_k\chi$. 
By replacing ${\rho_i}^j,_k$ in ${\cal H}^{(0,2)}$ according to (\ref{rsg})
the second order contribution to the perturbative expansion 
is readly evaluated by the formula (\ref{A1})  
\begin{eqnarray}
{\cal H}^{(0)}&\longrightarrow& 
\left(-\varepsilon^{ij}\Gamma_{ik}^k\omega_j
      +{1\over4}\Gamma_{ij}^k\Gamma_{ik}^j
      +{1\over2}\Gamma_{ij}^j\Gamma_{ik}^k
      -{3\over4}\Gamma_{ii}^j\Gamma_{jk}^k \right) J^2\nonumber\\
& &   +{3\over16}\Gamma_{ij}^k\Gamma_{ik}^j
      -{3\over16}\Gamma_{ii}^j\Gamma_{jk}^k 
\label{zero}
\end{eqnarray}
Quite surprisingly 
a term explicitly depending on $\omega_k$ survives.

\subsection*{Second order contribution from ${\cal H}^{(1)}$}

 The first order term of expansion (\ref{H2}) is cubic in the 
kinematical momenta, 
${\cal H}^{(1)}=-{1\over2}b^{1/2}(\partial_l\gamma^{ij})\varepsilon^{lk}
\bar{\Pi}_i\bar{\Pi}_k\bar{\Pi}_j$. As shown in the appendix,
this contribute the adiabatic expansion an ${l_B}^2$ order term
that may be directly evaluated by means of (\ref{A2}). 
The only necessary preparation  
is that of re-expressing $\partial_k\gamma^{ij}$ in terms of
Christoffel symbols. This is done by rewriting 
$\partial_\kappa\gamma_{\mu\nu}$ in terms of 
$\partial_\kappa\gamma^{\mu\nu}$ in the definition
$\Gamma_{\mu\nu}^\rho$
and by symmetrizing. The contraction with the 
zwei-beinen produces 
\begin{equation}
\partial_k\gamma^{ij}=-(\Gamma_{jk}^i+\Gamma_{ik}^j)
\label{dg=G}
\end{equation}
By substituting in (\ref{A2}) we obtain
\begin{eqnarray}
{\cal H}^{(1)}&\longrightarrow&
\left(-{3\over4}\Gamma_{ij}^k\Gamma_{ij}^k
      -{3\over4}\Gamma_{ij}^k\Gamma_{ik}^j
      -{3\over4}\Gamma_{ij}^j\Gamma_{ik}^k 
      +{3\over2}\Gamma_{ii}^j\Gamma_{jk}^k \right) J^2\nonumber\\
& &   -{3\over16}\Gamma_{ij}^k\Gamma_{ij}^k
      -{7\over16}\Gamma_{ij}^k\Gamma_{ik}^j
      +{1\over16}\Gamma_{ij}^j\Gamma_{ik}^k
      +{3\over8}\Gamma_{ii}^j\Gamma_{jk}^k
\label{one}
\end{eqnarray}

\subsection*{Second order contribution from ${\cal H}^{(2)}$}

A similar computation have to be performed for the second order 
term of the perturbative expansion, 
${\cal H}^{(2)}={1\over4}(\partial_m\partial_n\gamma^{ij})
\varepsilon^{mk}\varepsilon^{nl}\bar{\Pi}_i\bar{\Pi}_k\bar{\Pi}_l\bar{\Pi}_j$.
This time is necessary to re-express
the second order derivatives of the inverse metric in terms of
the Christoffel symbols and their derivatives. This is simply 
obtained by derivating (\ref{dg=G})
\begin{equation}
\partial_m\partial_n\gamma^{ij}=
-\partial_m\Gamma_{nj}^i-\partial_m\Gamma_{ni}^j
+\Gamma_{im}^h\Gamma_{hn}^j+\Gamma_{jm}^h\Gamma_{hn}^i
+\Gamma_{mh}^i\Gamma_{nh}^j+\Gamma_{mh}^j\Gamma_{nh}^i
\end{equation}
Recalling the definition of the scalar curvature 
$R=\partial_i\Gamma_{jj}^i-\partial_i\Gamma_{ij}^j
 +\Gamma_{ii}^j\Gamma_{jk}^k-\Gamma_{ij}^k\Gamma_{ik}^j$,  
formula (\ref{A1}) yields the second order
contribution produced by ${\cal H}^{(2)}$ 
\begin{eqnarray}
{\cal H}^{(2)}&\longrightarrow& 
\left( {1\over4}R
      -{1\over4}\partial_i\Gamma_{ij}^j
      +{3\over4}\Gamma_{ij}^k\Gamma_{ij}^k
      +{1\over2}\Gamma_{ij}^k\Gamma_{ik}^j
      -{1\over4}\Gamma_{ij}^j\Gamma_{ik}^k 
      -{1\over2}\Gamma_{ii}^j\Gamma_{jk}^k \right) J^2\nonumber\\
& &   -{1\over16}R
      -{5\over16}\partial_i\Gamma_{ij}^j
      +{3\over16}\Gamma_{ij}^k\Gamma_{ij}^k
      +{1\over4} \Gamma_{ij}^k\Gamma_{ik}^j
      +{1\over16}\Gamma_{ij}^j\Gamma_{ik}^k
      +{1\over8} \Gamma_{ii}^j\Gamma_{jk}^k
\label{two}
\end{eqnarray}
The two terms still containing derivatives of the Christoffel
symbols may be expressed in terms of derivatives of the magnetic field 
norm $b$ and contractions of the $\Gamma_{ij}^k$s simply by derivating
equation ({\ref{r1}), $\partial_i\Gamma_{ij}^j=-b^{-1}\triangle{b}
+b^{-2}|\nabla{b}|^2+\Gamma_{ii}^j\Gamma_{jk}^k$.  

\subsection*{}
 The effective Hamiltonian describing the motion of a charged 
particle to the second order in the adiabatic parameter ${l_B}$
is finally obtained by adding to $b(\bar{\Xi})J$ the  
contributions (\ref{zero}), (\ref{one}), (\ref{two}) as well as 
the term $-\nabla b/4b +|\nabla b|^2/8b^2$. As one 
have to be expect all the  contractions of the $\Gamma_{ij}^k$
but $\Gamma_{ij}^j\Gamma_{ik}^k$ cancel. This may be rewritten in
terms of $\nabla b$ by means of (\ref{r1}). We obtain
\begin{eqnarray}
{\cal H}&=& {b J\over{l_B}^2} +
    \left( {1\over4} R
                +{\nabla b\over b}\times\omega
                +{1\over4}{\triangle b\over b}
                -{3\over4}{|\nabla b|^2\over b^2}
          \right)J^2 \nonumber\\
        & & -{1\over16} R
            +{1\over16}{\triangle b\over b}
            -{1\over16}{|\nabla b|^2\over b^2}
            +{\cal O}({l_B})
\label{Heff}
\end{eqnarray}
All the functions are evaluated in the couple if conjugate 
operators $\bar{\Xi}^1$ and $\bar{\Xi}^2$. As before 
this expression has to be interpreted as the effective 
Hamiltonian describing the motion of the slow freedom while the 
particle is frozen in one of the $J$ eigenstates. The second
term is the correction surviving in the classical limit
while the remaining ones are of a pure quantal nature.

Even if our computation has been carried out in a Darboux 
coordinate frame, Eq.\ref{Heff} is explicitly covariant 
so that we are free to transform back to the original 
--arbitrary-- coordinates $x^\mu$. 
The price to pay is that of dealing with non-canonical
operators, the Hamiltonian getting evaluated in  
$X^\mu=x^\mu(\bar{\Xi})$. These `guiding center
variables' satisfy  in fact the non-canonical commutation relations  
$[X^2,X^1]=i{l_B}^2 b^{-1}(X)$ (compare references \cite{Lit,Mar}).  

 The effective dynamics is sensitive to the background scalar 
curvature. This coupling is particularly relevant when the magnetic 
two-form is proportional --in arbitrary coordinates-- to the 
volume two-form, $b_{\mu\nu}={l_B}^{-2}\sqrt{g}\varepsilon_{\mu\nu}$.
$g_{\mu\nu}$ and $b_{\mu\nu}$ define then a K\"ahler structure on 
$\Sigma$. The particle interacts only with the surface. The magnetic 
force becomes the `geometric gravitational' force of Cangemi and
Jackiw \cite{CJ}. In the strong magnetic regime the effective 
Hamiltonian driving the slow motion is proportional to the 
scalar curvature. In the semiclassical regime test 
particles drift along the line of constant 
curvature of the surface $\Sigma$.

 The effective dynamics is coupled tho the background spin-connection
as well. The coupling is not explicitly gauge invariant. A gauge 
transformation $\omega_k\rightarrow\omega_k+\partial_k\chi$
adds the term ${l_B}^2b^{-1}\varepsilon^{ij} (\partial_ib)(\partial_j\chi)
J^2$ to the Hamiltonian.
Gauge invariance may nevertheless be restored by the unitary transformation  
$U=e^{ib^{-1}J\chi}$. 
The second order term  $-ib^{-1}[b(\bar{\Xi}),\chi(\bar{\Xi})]J^2$ 
produced in this way brings (\ref{Heff}) in the original form.  

 Last but not the least, it is worth to mention that expansion 
(\ref{Heff}) yields the correct  flat limit \cite{Lit,Mar} supplying 
a full canonical derivation of it. 

\newpage
\section{Around a conical singularity}

 A typical situation of interest in 2+1 gravity is the motion around
a conical singularity \cite{DJtH}. As an example we consider therefore a 
charged particle on a conical surface subjected to an axisymmetric 
magnetic field 
decreasing as the inverse of the distance from the vertex. 
The problem is explicitly solvable allowing a check of our strong
magnetic field expansion. 

 The cone is parameterized by the distance $\rho$ from the vertex,
ranging $0\leq\rho\leq +\infty$, and the angle $\phi$, ranging
$0\leq\phi\leq2\pi$; the points $\phi=0$ and $\phi=2\pi$ are identified.
 In these coordinates metric and magnetic two-form take the form
\begin{equation}
g_{\mu\nu}=
\left(
\begin{array}{cc}
1 & 0 \\
0 & \alpha^2\rho^2 
\end{array}
\right)
 \ \ \ \ \ 
b_{\mu\nu}={1\over{l_B}^{2}}
\left(
\begin{array}{cc}
 0 & 1 \\
-1 & 0
\end{array}
\right)  
\end{equation}
where $\alpha$ is the conical angle; setting $\alpha=1$ brings the cone 
in the Euclidean plane. Though the curvature of $\gamma_{\mu\nu}$
vanishes, the geometry of the space is non trivial. The spin-connection
of the surface reads  $\omega_\mu=(0,\alpha)$ and can not be
gauged away for non-integer values of $\alpha$.

 We first consider the exact solution. In order to have a deeper
insight into the problem we focus on the classical motion, the 
discussion of quantum problem proceeding essentially along the same 
lines. Choosing the vector potential as $a_\mu=(0,{l_B}^{-2}\rho)$ the
Hamiltonian of the system writes 
\begin{equation}
{\cal H}={1\over2}{p_\rho}^2+
    {1\over2\alpha^2\rho^2}\left({p_\phi}-{\rho\over{l_B}^2}\right)^2  
\end{equation}
Given to the axial symmetry the momentum $p_\phi$ is conserved,
$\{{\cal H},p_\phi\}=0$, and can be replaced  by its constant value 
$L$. The radial motion of the system takes place in the 
effective Keplerian potential
\begin{equation}
V_{eff}(\rho)={L^2\over2\alpha^2\rho^2}
              -{L\over\alpha^2{l_B}^2\rho}
              +{1\over\alpha^2{l_B}^4}
\label{kepler}
\end{equation} 
where $L/\alpha^2{l_B}^2$ appears as an attractive Newton constant,
$L/\alpha$ as the angular momentum
and the whole spectrum in shifted 
by the energy $1/\alpha^2{l_B}^4$. The presence of the magnetic field 
produces bound states in the system.
There is no need to go through the well known solution of 
this problem, we focus instead on the qualitative behavior of the system 
in the strong magnetic regime. 
For small values of ${l_B}$ the minimum $\bar{\rho}=
L{l_B}^2$ of the effective potential becomes extremely deep and narrow.
 $V_{eff}$ is very well approximated by a 
harmonic oscillator centered in $\bar{\rho}$ and with frequency 
$\omega=1/\alpha L{l_B}^4$.
 While rotating around the axis of the cone at a distance $\bar{\rho}$
the particle performs 
therefore very rapid oscillations. The result
is that of a very thin and dense spiral wrapping around an 
orbit of constant radius. Neglecting the rapid oscillation,  
the effective angular velocity may be 
evaluated by eliminating $L$ in favor 
of $\bar{\rho}$ in the  relation
$L=p_{\phi}=\alpha^2\rho^2\dot{\phi}$. This yields  
\begin{equation}
\dot{\phi}\approx{1\over\alpha^2{l_B}^2\bar{\rho}}
\label{d}
\end{equation}    
The angular velocity distribution gives  
informations on the conical angle $\alpha$.

 We come now to the strong magnetic expansion (\ref{Heff}). 
Observe that the coordinates $\rho$ and $\phi$ are already of 
Darboux's type. The rapid oscillation of the particle have 
obviously to be identified with the freedom $\Pi_\rho-\Pi_\phi$ while 
the drift on the cone with the  motion of the
guiding center variables $R=\rho+{l_B}\Pi_\phi$ and 
$\Phi=\phi-{l_B}\Pi_\rho$. 
The coordinates $\rho$, $\phi$ and the couple of 
conjugate variables $R$, $\Phi$ parameterize two 
phase-space surfaces very close to each other and may 
be confused when order higher than ${l_B}$ are neglected.
The Hamiltonian driving the 
effective motion is immediately obtained from (\ref{Heff})
by evaluating gradient and Laplacian of the magnetic field 
norm $b(\rho)={1/(\alpha{l_B}^2\rho)}$;
the Galilei covariant components of the spin-connection are 
given by $\omega_i=(0,1/\rho)$;
\begin{equation}
{\cal H}_{eff}={1\over\alpha{l_B}^2R}J-{1\over2R^2}J^2+...
\label{HeffC}
\end{equation}
The angle $\Phi$ does not appear in the Hamiltonian so that $R$ 
is a second constant of motion besides $J$. 
The particle moves around the axis at the constant value of the 
radius $\bar{R}$.
The angle $\Phi$ evolves linearly in time according
to Hamilton equations $\dot{\Phi}=J/\alpha\bar{R}+
{\cal O}({l_B}^2)$. The adiabatic invariants $J$ and $R$ are
directly related by the conservation of energy. 
 Recalling that the classical system is in the adiabatic 
regime for small values of the total energy 
we  re-obtain an angular velocity distribution 
with the behavior (\ref{d}).

\section{Conclusions}

 The purpose of this paper was that of showing how is possible 
to set down a systematic canonical perturbative analysis for the 
motion of charged particles in a curved background geometry. 
This bridges the gap between the classical canonical theory and 
the non-canonical averaging methods traditionally employed in the 
classical analysis. Most important, the method allows
a direct discussion of the quantum case extending to this realm 
the whole classical `guiding center' picture. The aim is 
essentially achieved by means of Darboux transformations, 
standard averaging methods and elementary differential geometry.
For the shake of simplicity we restricted our attention to 
2+1 dimensions. Aside from its importance in the 
discussion of the whole 3+1 dimensional problem, the 2+1
dimensional system is already of a certain applicative 
importance in itself. An immediate application concerns the 
investigation of the non-minimal coupling of Cangemi and Jackiw 
in a Wick-rotated two dimensional gravity.
More in general, Hamiltonian (\ref{Heff}) gives us
immediate informations on how wavefunctions and eigenvalues 
of an electron in a quantum-Hall like device are modified 
when a small inhomogeneity of the  magnetic field 
or of the thin film geometry are introduced.
The electron behaves like a one degree of freedom system having 
the thin film --the spatial surface $\Sigma$-- as `phase space'. 
The fast freedom is still frozen in a harmonic oscillator eigenstate
and the discussion of section III indicates how the harmonic oscillator
eigenfunctions have to be constructed.
 The peculiar way the slow freedom  couples to the `phase space' 
scalar curvature and spin-connection is particularly intriguing 
and deserve further investigation. Another important issue concern 
the convergence of the perturbative expansion, which is expected 
to be in general an asymptotic series.
 We conclude by pointing out that considering more spatial dimensions produces
other interesting phenomena --the coupling of the effective dynamics 
of the new freedoms with geometry induced gauge structures-- that
can be described essentially by the same formalism. The inclusion of 
spin is also quite immediate. 
The restriction to 2+1 dimensions allowed us to single out the
effective coupling with the background
geometry without mixing it with phenomena of a different nature. 
The effective motion in 3+1 dimensions and the inclusion of 
spin will be considered in future publications.

\section*{Acknowledgments}

I am grateful to R. Jackiw for helpful discussions and comments.

\newpage
\appendix
\section*{Averaging around a harmonic oscillator}

  The introduction of a suitable set of variables reduces 
the study of the effective motion of a charged particle  
to the discussion of a Hamiltonian of the form
\begin{eqnarray}
{\cal H}=\alpha J + \epsilon   \alpha^{ijk} \Pi_i\Pi_k\Pi_j
                  + \epsilon^2 \alpha^{ijkl}\Pi_i\Pi_k\Pi_l\Pi_j
                  + ...
\nonumber
\end{eqnarray}
As in the body of the paper $J={1\over2}({\Pi_1}^2+{\Pi_2}^2)$,
and $\Pi_1$, $\Pi_2$ are a couple of conjugate variables,
$[\Pi_1,\Pi_2]=i$. $\epsilon$ is a small parameter. 
 The coefficients appearing in the expansion
are allowed to depend on slow variables.
The self-adjointnes of ${\cal H}$ requires $\alpha^{ijk}=\alpha^{jik}$
and $\alpha^{ijkl}=\alpha^{jilk}$.
 For a charged particle in a strong magnetic field the coefficients 
$\alpha$, $\alpha^{ijk}$, $\alpha^{ijkl}$, ..., are quite complicated 
expressions involving the spin-connection, the metric tensor and 
their derivatives evaluated in the slow guiding center variables 
$\Xi^1$ and $\Xi^2$, $[\Xi^2,\Xi^1]=i\epsilon^2$.
Very handy formulas will be worked out in this appendix in order to 
maximally simplify the manipulation of this expressions.

 As far as $\epsilon$ is set equal to zero 
dynamics is described by $h^{(0)}\equiv\alpha J$. The system
behaves as a harmonic oscillator with frequency depending
on the non-dynamical parameters $\Xi^i$. A nonzero value of 
$\epsilon$ turns the perturbation on, making, on the same 
time, the guiding center operators in a couple of conjugate 
dynamical variables. 
 In order to extract the effective dynamical content of the 
theory to the various order in the perturbative parameter 
$\epsilon$ we will subject the system to a series of near-identity 
unitary transformations. These are chosen in such a way 
that the various terms of the perturbative expansion 
depend on $\Pi_1$ and $\Pi_2$ only though $J$ and it powers.
 This makes $J$ into an adiabatic invariant ---a quantity 
conserved up to higher order of some power of $\epsilon$--- 
and  allows to identify the Hamiltonian driving the effective 
motion of the slow variables in correspondence of every value 
taken by $J$. The technique is based  essentially  on the identity
$\mbox{e}^{ia}{\cal H}\mbox{e}^{-ia}={\cal H}+ i[a,{\cal H}]
-{1\over2}[a,[a,{\cal H}]]+...$, where
$a=1+\epsilon a^{(1)}+\epsilon^2a^{(2)}+...$ is the generator 
of a nearly-identity unitary transformation.
 The self-adjoint operators $a^{(1)}$, $a^{(2)}$, etc. have to be chosen
order by order in such a way that the desired conditions are matched.

 We start by the order $\epsilon$ of the expansion:
$h^{(1)}\equiv\alpha^{ijk} \Pi_i\Pi_k\Pi_j$. Note that 
since $\Pi_i\Pi_k\Pi_j+\Pi_j\Pi_k\Pi_i$ is completely symmetric in
the indices $i$, $j$ and $k$ only the completely symmetric part
of $\alpha^{ijk}$ matters. We can therefore assume the complete symmetry
of the $\alpha^{ijk}$.
It is then easy to verify that the choice
\begin{eqnarray}
a^{(1)}=-{1\over3}\alpha^{-1}
\left(\alpha^{ijl}+2\delta^{ij}\alpha^{hhl}\right)
\varepsilon_{lk}\Pi_i\Pi_k\Pi_j  
\nonumber
\end{eqnarray}
produces the  counterterm $i[a^{(1)},h^{(0)}]=-h^{(1)}$.
The first order term of the transformed  expansion 
vanishes identically. The operation is nevertheless not painless. 
The transformation contribute in fact the second order term 
$h^{(1,2)}={i\over2}[a^{(1)},h^{(1)}]$. This can be evaluated in
\begin{eqnarray}
h^{(1,2)}&=&{3\over2}\left(
                      {\alpha^{ihh}\alpha^{jkl}\over\alpha}
                     +{\alpha^{jhh}\alpha^{ikl}\over\alpha}
                     -{\alpha^{ijh}\alpha^{klh}\over\alpha}
                     -2{\delta^{ij}\alpha^{hhg}\alpha^{klg}\over\alpha}
                     \right)\Pi_i\Pi_k\Pi_l\Pi_j\nonumber\\
& &-{\alpha^{ijk}\alpha^{ijk}\over\alpha}
   +3{\alpha^{iik}\alpha^{jjk}\over\alpha}
\nonumber
\end{eqnarray}
 The problem is re-conduced to the discussion of the second order term.

  Focus therefore on $h^{(2)}\equiv\alpha^{ijkl}\Pi_i\Pi_k\Pi_l\Pi_j$.
The symmetrization in the various couples of indices can still be
performed producing contributions of the form $\alpha^{ijkl}
\varepsilon_{ij}\varepsilon_{kl}$ etc.,  not depending on 
$\Pi_1$ and $\Pi_2$. Nevertheless a quite handy expression 
can already be obtained by assuming only the symmetrization of the first
and second couples of indices; the case we have to deal with.
A few computation show then that the right choice to make 
the second order of the perturbative expansion to depend only
on  powers of $J$ is  
\begin{eqnarray}
a^{(2)}= -{1\over8}\alpha^{-1}\left[
          \alpha^{ijkh}+
          \delta^{ij}\left(\alpha^{kghg}+{1\over2}\alpha^{khgg}\right)\right]
          \varepsilon^{hl}\Pi_i\{\Pi_k,\Pi_l\}\Pi_j
\nonumber
\end{eqnarray}
The second order term $i[a^{(2)},h^{(0)}]$ produced by this
transformation combines with $h^{(2)}$ in such a way to give 
the final contribution to the perturbative expansion
\begin{equation}
\alpha^{ijkl}\Pi_i\Pi_k\Pi_l\Pi_j \longrightarrow
 \left(\alpha^{ijij}+{1\over2}\alpha^{iijj}\right)J^2 
-{1\over4}\alpha^{ijij} + {5\over8}\alpha^{iijj}
\label{A1}
\end{equation}
No matter how complicated is $h^{(2)}$, 
formula \ref{A1} allows to immediately write down the contribution 
to the  effective dynamics by evaluating a few contractions of the
coefficients $\alpha^{ijkl}$. The first 
application of \ref{A1} is  the second order
contribution produced by $h^{(1)}$ through $h^{(1,2)}$. 
A brief computation yields the quite compact formula
\begin{equation}
\alpha^{ijk}\Pi_i\Pi_k\Pi_j \longrightarrow
-\left({3\over2}{\alpha^{ijk}\alpha^{ijk}\over\alpha}+
       {9\over4}{\alpha^{iik}\alpha^{jjk}\over\alpha}\right)J^2
-{5\over8}{\alpha^{ijk}\alpha^{ijk}\over\alpha}
+{3\over16}{\alpha^{iik}\alpha^{jjk}\over\alpha}
\label{A2}
\end{equation}
Again, the contribution to the effective dynamics produced by 
$h^{(1)}$ may be obtained through \ref{A2} by evaluating a
few contraction on the square of the coefficients $\alpha^{ijk}$.

\end{document}